\documentclass[amsmath,amssymb,
aps,
nofootinbib,
prd,
superscriptaddress,
tightenlines,
notitlepage,
twocolumn,
showpacs,
floatfix]{revtex4-2}
\usepackage[utf8]{inputenc}
\usepackage{color}
\usepackage{xcolor}
\usepackage{hyperref}
\usepackage{graphicx}% Include figure files
\usepackage{dcolumn}% Align table columns on decimal point
\usepackage{bm}% bold math
\newcommand{\be}{\begin{equation}}
\newcommand{\ee}{\end{equation}}
\newcommand{\bea}{\begin{equation}\begin{aligned}}
\newcommand{\eea}{\end{aligned}\end{equation}}

\newcommand{\IHEP}{\affiliation{Key Laboratory of Particle Astrophysics, Institute of High Energy Physics,
Chinese Academy of Sciences, Beijing 100049, China}}

\begin{document}

\preprint{APS/123-QED}

\title{
Axion Minicluster Halo Limits from Wide Binary Disruption}
\author{Zihang Wang}\email{wangzihang@ihep.ac.cn}\IHEP
\author{Yu Gao}\email{gaoyu@ihep.ac.cn}\IHEP
\date{\today}
\begin{abstract}
Axionic dark matter can form miniclusters and minicluster halos from inhomogenuities in the early Universe. If MCHs are sufficiently massive, their existence can be revealed by small-scale gravitational tidal perturbation to halo-like binary star system in the Galaxy. The observed population of the Milky Way's wide-separation binaries with $a\gtrsim\mathcal{O}(0.1)\,$ parsec offer a sensitive test to dynamic evaporation from MCHs. Considering data from recent GAIA observations, we derive significant constraints on the MCH fraction of the galactic dark matter halo. For several scenarios including dense MCHs and isolated minicluster models, these limits will apply to axion-like particles in the mass range $m_{a}\sim 10^{-15}-10^{-12}\,\rm eV$  and $m_{a}\sim 10^{-19}-10^{-16}\,\rm eV$, respectively.
\end{abstract}

\maketitle

\section{Introduction}
Numerous astrophysical observations show the majority the matter in the Universe is in the form of dark matter~\cite{Planck:2018vyg}. The axion, originally proposed to solve the strong CP problem~\cite{Peccei:1977hh,Weinberg:1977ma,Wilczek:1977pj}, is also one of the best motivated candidates for dark matter~\cite{Sikivie:1982qv,Preskill:1982cy}. High-energy breaking of the global U(1) Peccei–Quinn (PQ) symmetry, leaves a Goldstone boson that acquires a mass near the QCD phase transition scale, and naturally relaxes the CP-violating angle $\bar{\theta}$ to zero. The low energy potential of the axion field makes its late-time energy density evolution matter-like~\cite{Turner:1983he}. More general axion-like particles (ALPs) are also motivated, for example by dimension compatification in string theory~\cite{Svrcek:2006yi} and many other beyond standard model theories~\cite{DiLuzio:2020wdo}. 

The axion's energy density is typically produced through the `misalignment' mechanism and from the decay of topological defects, see Ref.~\cite{Marsh:2015xka} for review. When $U(1)_{PQ}$ symmetry is broken after the end of inflation, the axion field is expected to vary from one horizon to another. Such primordial fluctuations will grow as the Universe expands, and form dense regions a.k.a. miniclusters (MCs)~\cite{Kolb:1994fi,Chang:1998tb}. At later times, axion MCs may further aggregate into larger structures called minicluster halos (MCHs)~\cite{Fairbairn:2017sil}. A significant part of today's axion dark matter can be contained within MCs or MCHs. 

The formation of axion MCs and MCHs, and the detection of these structures are crucial topics. Semi-analytical calculations~\cite{Fairbairn:2017sil} and numerical simulations~\cite{Eggemeier:2019khm,Xiao:2021nkb,Ellis:2022grh} 
could provide the mass function of MCHs. Recent studies have proposed that axion MCs and MCHs can be detectable through microlensing~\cite{Fairbairn:2017sil}. The gravitational effects of MCHs may also be detectable by pulsar timing arrays~\cite{Siegel:2007fz,Lee:2020wfn}, as well as by observing the dynamic heating of sizeable systems. For example, see recent studies on the relaxation effect by dwarf-galaxy sized axion halos ~\cite{Bar-Or:2018pxz,2019ApJ...885..155W,2020PrPNP.11303787N,Church:2018sro}. Axion MC and MCHs can vary in size and lead to gravitation fluctuation on their characteristic scales. Similar to soliton case~\cite{Marsh:2018zyw}, spatially extended MCHs can surely contribute to a relaxation process. By studying the dynamical heating of stars by axion MCs and axion stars in ultrafaint dwarfs, the mass of the axion can be constrained~\cite{Chang:2024fol}.

In this work we consider binary star disruption form short-scale gravitational perturbation from axion MCs and MCHs. Wide binary star systems are numerous in our Galaxy, and their maximal separation can be larger than $0.1$ parsec~\cite{Tian:2020ApJS}. This is still a much smaller scale compared to the size of star clusters and dwarfs. Tidal effects on this scale can potential reveal a MC and MCH population with sub-parsec sizes, and correspondingly a different axion particle mass. Recently, binary star disruption by scale-dependent tidal disruption has been studied analytically~\cite{Qiu:2024muo}. Stochastic tidal perturbation increases the orbital energy and evaporates weakly bound binary systems, leading to potential reduction in the number of binary stars with relevant sizes that can be tested against galactic star survey observations~\cite{El-Badry:2021MNRAS,Klioner:2021AA}.

This paper is organized as follows: in Section~\ref{sec:a}, we present the MCH mass functions from previous studies. In Section~\ref{sec:b}, we calculate the evaporation rate of binary stars. In Section~\ref{sec:c}, we consider different models of MCHs and establish constraints for ultralight axions. For convenience, we will use the word `axion' to denote axion-like particle in the discussion of their tidal effects, unless otherwise specified.

%Ultralight axions recently becomes popular because they may help explain the small scale crisis faced by cold dark matter theories. 

\section{Axion minicluster halo mass functions}
\label{sec:a}

We briefly recast the analytical derivation of MCH mass function here. Different models will be considered with their specific mass function and concentration.

In the early Universe, the axion field $\varphi$ evolves as
\begin{equation}
\ddot{\varphi}+3H(t)\dot{\varphi}+m_{a}^{2}(t)\varphi=0\, ,
\end{equation}
where $m_{a}(t)$ is the mass that can evolve with temperature~\cite{GrillidiCortona:2015jxo}. At time $t_{\rm osc}$, the axion field overcomes Hubble friction and starts to oscillate, determined by $m_{a}(t_{\rm osc})=3H(t_{\rm osc})$. The horizon size at $t_{\rm osc}$ sets the scale for the majority of axion field fluctuations. The corresponding wavenumber is $k_{\rm osc}\equiv a(t_{\rm osc})H(t_{\rm osc})$, and correlation diminishes beyond this scale. The axion overdensity is $\delta_{a}(\vec{x},t)=(\rho_{a}(\vec{x},t)-\bar{\rho}_{a}(t))/\bar{\rho}_{a}(t)$, and its Fourier transformation is
\begin{equation}\label{eq:ovd}
\delta_{\vec{k}}(t)=\int d^{3}\vec{x}\, \delta_{a}(\vec{x},t)e^{-i\vec{k}\cdot\vec{x}} \, ,
\end{equation}
The power spectrum $P_{k}$ is defined as,
\begin{equation}\label{eq:pow}
\langle\delta_{\vec{k}}(t)\delta_{\vec{k}'}(t)\rangle=(2\pi)^{3}P_{k}(t)\delta^{3}(\vec{k}-\vec{k'}) \, ,
\end{equation}

The initial power spectrum is often assumed to be white noise cutoff at $k_{\rm osc}$, namely
\begin{equation}\label{eq:pow2}
\Delta_{k}^{2}(t_{\rm osc})\equiv\frac{k^{3}}{2\pi^{2}}P_{k}(t_{\rm osc})=A_{\rm osc}\left(\frac{k}{k_{\rm osc}}\right)^{3}\Theta(k_{\rm osc}-k)\, ,
\end{equation}
where $A_{\rm osc}$ is a constant number of order unity. By considering misalignment axions with initial misalignment angle $\theta_{i}$ randomly varies from one horizon to another, $A_{\rm osc}$ is taken to be $12/5$~\cite{Fairbairn:2017sil}. In the QCD axion case, simulations yield $A_{\rm osc}\sim 0.01-0.3$~\cite{Vaquero:2018tib}. For ALPs, simulation in Ref.~\cite{OHare:2021zrq} gives $A_{\rm osc}=0.3$. The mass $m_{0}$ of a typical MC is characterised by the mass contained inside the Hubble horizon at $t_{\rm osc}$,
\begin{equation}\label{eq:m0}
m_{0}=\frac{4\pi^4}{3}\frac{\bar{\rho}_{a}(t_{\rm osc})}{H^{3}(t_{\rm osc})}=\frac{4\pi^4}{3}\frac{\bar{\rho}_{a0}}{k_{\rm osc}^{3}}\, ,
\end{equation}
where $\bar{\rho}_{a0}$ is the axion average density at present. 
As we will see below, $m_{0}$ also provides a characteristic mass scale for MCs and MCHs.

\begin{figure}[htbp]
  \centering
  \includegraphics[width=9cm]{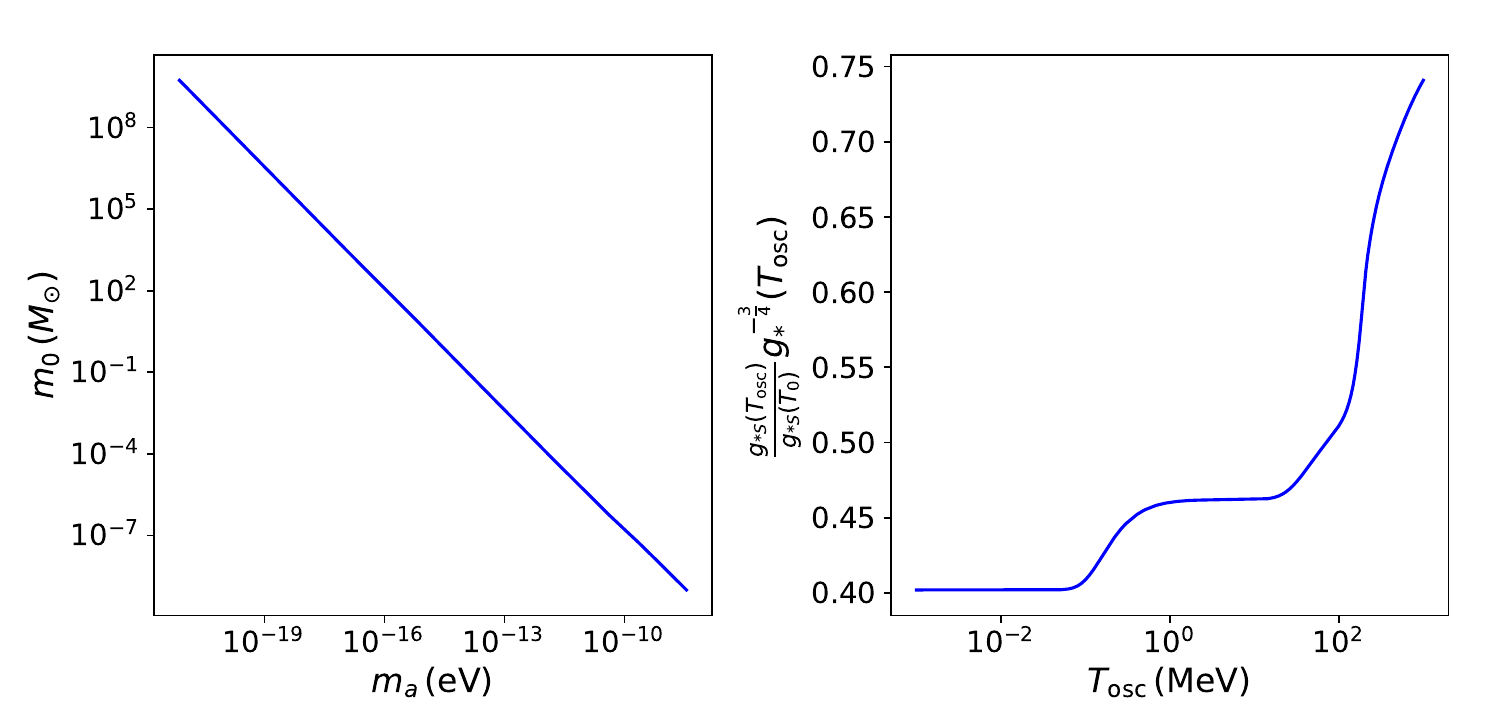}\\
  \caption{\emph{Left}: The MC characteristic mass $m_{0}$ for different axion mass. Here we ignore the temperature evolution of the ALP mass. \emph{Right}: The factor $[g_{*S}(T_{\rm osc})/g_{*S}(T_{0})]g_{*}^{-\frac{3}{4}}(T_{\rm osc})$ for different oscillating temperature $T_{\rm osc}$.}
  \label{img0}
\end{figure}

For ALPs from the string theory, its mass can be taken as temperature independent. In such case,
\begin{equation}\label{eq:M0}
m_{0}=281M_{\odot}\left(\frac{m_{a}}{10^{-16}\,{\rm eV}}\right)^{-\frac{3}{2}}\frac{g_{*S}(T_{\rm osc})}{g_{*S}(T_{0})}g_{*}^{-\frac{3}{4}}(T_{\rm osc})\, ,
\end{equation}
where $T_{\rm osc}$ and $T_{0}$ are the temperature at $t_{\rm osc}$ and the cosmic microwave background temperature at present. $g_{*}$ and $g_{*S}$ are effective relativistic degrees of freedom and effective entropy degrees of freedom, whose values are taken from Ref.~\cite{Laine:2015kra}. The factor on the right hand side of Eq.~(\ref{eq:M0}) is of order unity. From Eq.~(\ref{eq:M0}) we can see that the MC mass is larger for smaller axion mass.
In the left panel of Fig.~\ref{img0} we plot the MC mass for different temperature independent axion mass. In the right panel of Fig.~\ref{img0}, the factor $[g_{*S}(T_{\rm osc})/g_{*S}(T_{0})]g_{*}^{-\frac{3}{4}}(T_{\rm osc})$ is plotted for different oscillating temperature, which is close to unity.

The subhorizon perturbations do not grow significantly until matter-radiation equity, hence $\Delta_{k}(t_{\rm eq})\approx\Delta_{k}(t_{\rm osc})$.
The variance of density fluctuation is given by

\begin{equation}\label{eq:sigma}
\sigma^{2}\equiv\left(\frac{\delta m_{s}}{m_{s}}\right)^{2} =\frac{\left\langle\left[\int d^{3}\vec{x}\,\delta_{a}(\vec{x})W(\vec{x})\right]^{2}\right\rangle}{\left[\int d^{3}\vec{x}\, W(\vec{x})\right]^{2}}\, ,
\end{equation}
where a window function $W(\vec{x})$ is introduced. We choose a Gaussian window function $W(\vec{x})=e^{-r^{2}/(2R^{2})}$ with $r\equiv|\vec{x}|$, which corresponds to a comoving radius $R$. When considering MCH formation from the density fluctuations, $R$ and $m_{s}$ represent the comoving radius and the mass of MCHs. Using Eq.~(\ref{eq:ovd}),(\ref{eq:pow}) and (\ref{eq:pow2}), the variance can be written as
\begin{equation}\label{eq:sig2}
\sigma^{2}(R)=\int \frac{{\rm d}k}{k}\Delta^{2}(k)|\tilde{W}(kR)|^{2}\, ,
\end{equation}
where $\tilde{W}(kR)=e^{-\frac{1}{2}k^{2}R^{2}}$ is the Fourier transformation of real space window function $W(\vec{x})=e^{-r^{2}/(2R^{2})}$.

The window function chosen here provides a cutoff at large radius $r$, $\delta_{a}(r)\rightarrow \delta_{a}(r)e^{-r^{2}/(2R^{2})}$. The corresponding dark matter mass in the denominator of Eq.~(\ref{eq:sigma}) is
\begin{equation}\label{eq:msR}
m_{s}=(2\pi)^{\frac{3}{2}}\bar{\rho}_{a0}R^{3}\, .
\end{equation}

When considering structure growth in matter-dominated era, we must compare the modes with the Jeans scale~\cite{Guth:2014hsa,Wang:2020zur},
\begin{equation}
k_{J}=\left(16\pi Ga\bar{\rho}_{a0}m_{a}^{2}\right)^{\frac{1}{4}}\, .
\end{equation}
Both $k_{J}$ and $k_{\rm osc}$ are proportional to $m_{a}^{\frac{1}{2}}$. We can verify directly that $k_{\rm osc}<k_{J}(t_{\rm eq})$ independent of the ALP mass. 
For $k_{\rm osc}<k_{J}(t_{\rm eq})$, the growth of structures is scale independent after $t_{\rm eq}$, and linear structure growth leads to the growth factor,~\cite{Percival:2005vm}
\begin{equation}
D(a)=\frac{5\Omega_{m}}{2}E(a)\int_{0}^{a}\frac{{\rm d}a'}{[a'E(a')]^{3}}\, ,
\end{equation}
where $E^{2}(a)\equiv \Omega_{m}a^{-3}+\Omega_{k}a^{-2}+\Omega_{\Lambda}$. The growth factor gives the growth of perturbations after the matter-radiation equity. 
The variance of density fluctuations at present is given by
\begin{equation}\label{eq:sigma0}
\sigma_{0}=\frac{D(a_{0})}{D(a_{\rm eq})}\sigma_{\rm eq}\, ,
\end{equation}
and it is related to the halo mass function via the density collapse process, e.g., described by the Press-Schechter (PS) formalism~\cite{Press:1973iz} or the Sheth–Tormen (ST) formalism~\cite{Sheth:1999mn}, the latter including non-spherical collapses. We will consider both scenarios in this paper. In the 
PS case, the halo mass function is
\begin{equation}\label{eq:PS}
\frac{{\rm d}n}{{\rm d}\ln m_{s}}=\sqrt{\frac{2}{\pi}}\frac{\rho_{a0}}{m_{s}}\nu e^{-\frac{1}{2}\nu^{2}}\left|\frac{{\rm d}\ln\sigma}{{\rm d}\ln m_{s}}\right|\, ,
\end{equation}
where $\nu=\delta_{c}/\sigma(m_{s})$, $\delta_{c}=1.686$ is the critical overdensity for gravitational collapse~\cite{Ellis:2020gtq}. The ST formalism gives
\begin{equation}\label{eq:ST}
\frac{{\rm d}n}{{\rm d}\ln m_{s}}=A\sqrt{\frac{2}{\pi}}\frac{\rho_{a0}}{m_{s}}\sqrt{q}\nu \left[1+(\sqrt{q}\nu)^{-2p}\right]e^{-\frac{1}{2}q\nu^{2}}\left|\frac{{\rm d}\ln\sigma}{{\rm d}\ln m_{s}}\right|\, ,
\end{equation}
where the parameters are, $A=0.3222,\,p=0.3,\,q=0.707$~\cite{Sheth:1999mn}. When discussing MCHs in the solar neighborhood, $\rho_{a0}$ should also be taken as the local axion density. 

\begin{figure}[htbp]
  \centering
  \includegraphics[width=8cm]{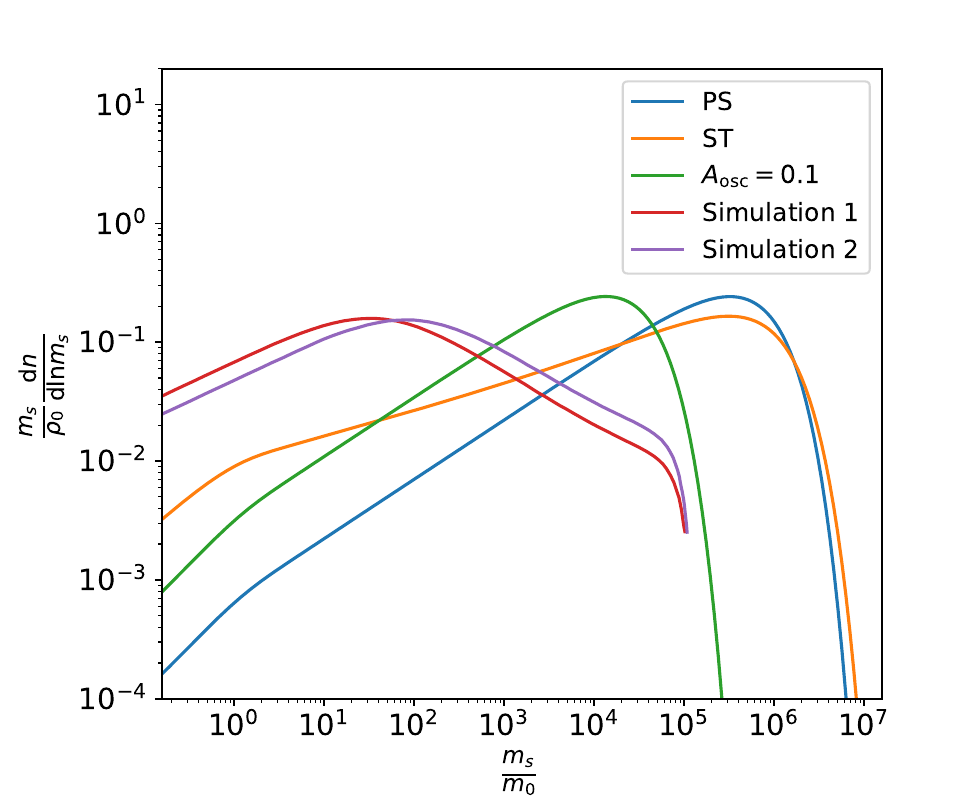}\\
  \caption{The MCH mass function for PS, ST formalism with different $A_{\rm osc}$. We also adopt simulation results of MCH mass function in Ref.~\cite{Xiao:2021nkb}.}
  \label{img1}
\end{figure}

MCs and MCHs reside in galactic halos, and their sizes are much smaller than galactic halos. The average density of axions within MCHs $\rho_0$ must be smaller than or equal to the local dark matter density $\rho_{\rm DM}$. 
The mass function of MCHs is normalized so that 
\begin{equation}\label{eq:norm}
\int \frac{{\rm d}n}{{\rm d}\ln m_{s}} {\rm d}m_{s}=\rho_{0} \, ,
\end{equation}
Here we emphasize the difference between the densities $\rho_{0}$, $\rho_{a0}$ and $\rho_{\rm DM}$. $\rho_{0}$ does not include axions outside MCHs. In contrast, $\rho_{a0}$ is the local axion density including those outside MCHs. Compared with $\rho_{a0}$, $\rho_{\rm DM}=0.4\,{\rm GeV/cm^{3}}$ includes other possible components of dark matter beside axions. 

We also introduce a factor $f_{\rm mc}$ defined as $\rho_{0}=f_{\rm mc}\rho_{\rm DM}$.
$f_{\rm mc}$ represents the fraction of the axions within MCHs. The other parts of the dark matter are assumed to be locally uniform and do not contribute to binary evaporation. A value of $f_{\rm mc}>1$ indicates the density of axions within MCHs exceeds the local dark matter density, hence the corresponding axion parameter is unconstrained.  

The prefactor on the right hand side of Eq.~(\ref{eq:PS}) and (\ref{eq:ST}) will be modified to satisfy the normalization Eq.~(\ref{eq:norm}). We normalize all the MCH mass functions separately. For example, for the PS formalism, we use the expression,
\begin{equation}\label{eq:PS2}
\frac{{\rm d}n}{{\rm d}\ln m_{s}}=\frac{C_{\rm PS}}{m_{s}}\nu e^{-\frac{1}{2}\nu^{2}}\left|\frac{{\rm d}\ln\sigma}{{\rm d}\ln m_{s}}\right|\, ,
\end{equation}
where a constant $C_{\rm PS}$ is introduced to satisfy Eq.~(\ref{eq:norm}). Hence, the mass function now depends on $\rho_{0}$ rather than $\rho_{a0}$. Physically, the original PS and ST formalism predict the value of $f_{\rm mc}$ (which is close to unity) if axions compose all the dark matter. But in this paper we treat $f_{\rm mc}$ as a free parameter. We only use the shape of the mass function from the PS and ST formalism while the normalization is fixed by Eq.~(\ref{eq:norm}).

Note that when $A_{\rm osc}$ is fixed, the MCH mass function $(m_{s}/\rho_{0}){\rm d}n/{\rm d}\ln m_{s}$ depends solely on $m_{s}/m_{0}$. This can be seen from Eq.~(\ref{eq:sig2}). Using Eq.~(\ref{eq:pow2}), the mass variance Eq.~(\ref{eq:sig2}) at $t_{\rm eq}$ can be rewritten as

\begin{equation}\label{eq:sig3}
\sigma^{2}_{\rm eq}(R)=\int \frac{{\rm d}(kR)}{kR}A_{\rm osc}\left(\frac{kR}{k_{\rm osc}R}\right)^{3}\Theta(k_{\rm osc}-k)|\tilde{W}(kR)|^{2}\, ,
\end{equation}

Hence $\sigma_{\rm eq}^{2}$ at $t_{\rm eq}$ only depends on $k_{\rm osc}R$ for fixed $A_{\rm osc}$. Using Eq.~(\ref{eq:m0}) and (\ref{eq:msR}) we find that $k_{\rm osc}R$ is directly related to $m_{s}/m_{0}$, namely
\begin{equation}
\frac{m_{s}}{m_{0}}=\frac{3(2\pi)^{\frac{3}{2}}}{4\pi^{4}}(k_{\rm osc}R)^{3}\, .
\end{equation}
The mass variance at present $\sigma_{0}$ is related to $\sigma_{\rm eq}$ through Eq.~(\ref{eq:sigma0}), where the factor $D(a)$ only depends on cosmological parameters.
Hence $\nu$ and ${\rm d}\ln\sigma/{\rm d}\ln m_{s}$ only depends on $m_{s}/m_{0}$. The PS and ST formalisms Eq.~(\ref{eq:PS}) and (\ref{eq:ST}) shows that (after normalization procedure)
$(m_{s}/\rho_{0}){\rm d}n/{\rm d}\ln m_{s}$ depends solely on $m_{s}/m_{0}$. The shape of MCH mass function is independent of $m_{0}$. Changing ALP mass only alters $m_{0}$, not the mass function $(m_{s}/\rho_{0}){\rm d}n/{\rm d}\ln m_{s}$. In the following, we use this property to extrapolate the simulation results in Ref.~\cite{Xiao:2021nkb} to different axion masses.

In Fig.~\ref{img1}, we plot the normalized MCH mass function. The blue and orange lines show the MCH mass function from PS and ST formalisms with $A_{\rm osc}=12/5$. The green line is the MCH mass function from PS formalism with $A_{\rm osc}=0.1$. The red and purple lines are MCH mass functions obtained in simulation in Ref.~\cite{Xiao:2021nkb}. We denote the two MCH mass functions as simulation 1 and simulation 2, which correspond to different assumptions for termination of axion MCH growth.

Beside the halo mass function, the radius of MCHs is also important when considering its effects on binary stars. The density profile of MCHs can be described by Navarro-Frenk-White (NFW) profile~\cite{Navarro:1995iw},

\begin{equation}\label{eq:NFW}
\rho(r)=\frac{\rho_{c}\delta_{\rm char}}{\frac{r}{r_{s}}\left(1+\frac{r}{r_{s}}\right)^{2}} \, .
\end{equation}
where $\rho_{c}$ is the critical density, $r_{s}$ is the scale radius of MCHs. The virial radius $r_{v}$ is defined as the radius where the average density of the dark matter is $200\rho_{c}$, namely
\begin{equation}\label{eq:rv}
r_{v}=\left(\frac{3}{800}\frac{m_{s}}{\pi\rho_{c}}\right)^{\frac{1}{3}} \, ,
\end{equation}
Hence $r_{v}$ is directly related to the total mass of MCH. The concentration $c$ is defined as $c=r_{v}/r_{s}$. Using density profile Eq.~(\ref{eq:NFW}), we obtain a relation between $\delta_{\rm char}$ and $c$,

\begin{equation}\label{eq:cdelta}
\delta_{\rm char}=\frac{200}{3}\frac{c^{3}}{\ln(1+c)-\frac{c}{1+c}} \, ,
\end{equation}
The scale radius $r_{s}$ of MCHs can be obtained if we know the the mass $m_{s}$ and concentration $c$.
In principle, the concentration of MCHs has large uncertainties.
For the analytical MCH formation model presented here, we only consider two limiting cases~\cite{Fairbairn:2017sil}. Diffuse MCH model assumes that during MCH formation, the MCs merge and completely disrupt. Dense MCH model assumes that MCs tightly group together to form MCHs without change of individual MC density. For diffuse and dense MCHs, the mass functions are the same while dense MCHs have a smaller radius compared with diffuse MCHs with the same mass.  For the two different models considered here, we plot the concentration with respect to MCH mass. 

The MCH simulations (for both simulation 1 and 2) in Ref.~\cite{Xiao:2021nkb} gives the MCH concentration at $z=0$,

\begin{equation}
c=\frac{9.46\times 10^{3}}{\sqrt{m_{s}/(A_{\rm osc}m_{0})}} \, ,
\end{equation}
Note that the slight difference from Ref.~\cite{Xiao:2021nkb} in the numerator is due to the different definition of virial radius $r_{v}$.

\begin{figure}[htbp]
  \centering
  \includegraphics[width=8cm]{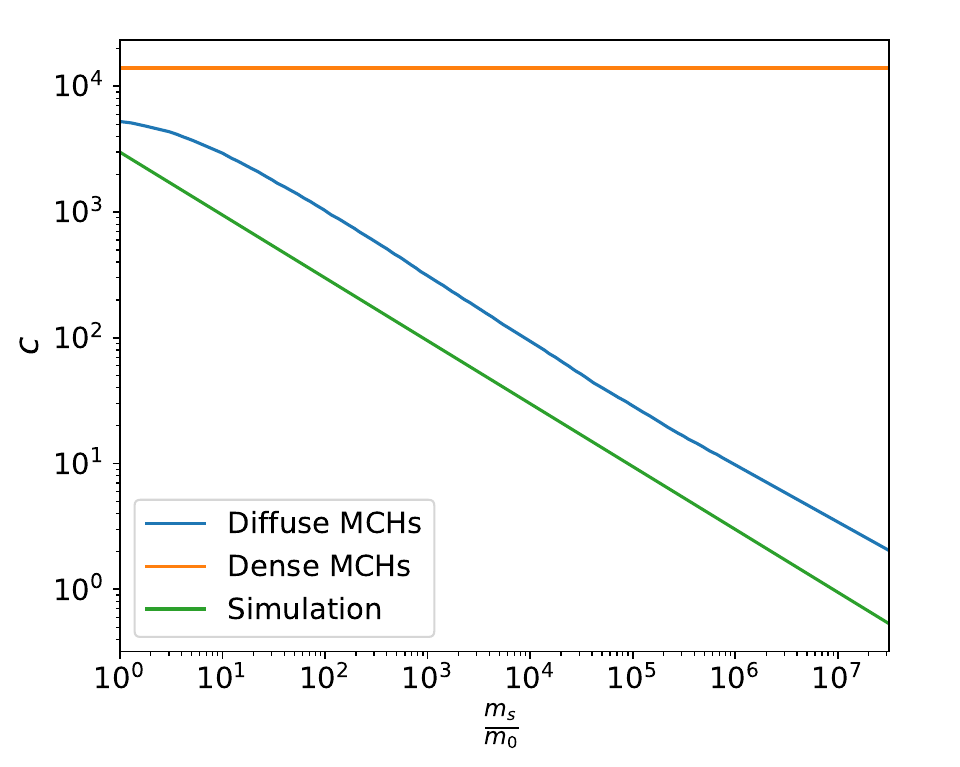}\\
  \caption{The MCH concentration we used for diffuse MCHs~\cite{Bullock:1999he,Fairbairn:2017sil}, dense MCHs and simulations in Ref.~\cite{Xiao:2021nkb}.}
  \label{img2}
\end{figure}

We will also consider the case that MCs are isolated~\cite{Kolb:1994fi,Fairbairn:2017sil}, in which MCs are spatially randomly distributed, but do not group together to form MCHs. These MCs have the same mass $m_{0}$, which is determined from the horizon size and axion density at $t_{\rm osc}$. 
The density of MC is~\cite{Kolb:1994fi,Kolb:1995bu}
\begin{equation}\label{eq:MCd}
\rho_{\rm mc}=140\delta^{3}(1+\delta)\bar{\rho}_{a}(z_{\rm eq})\, ,
\end{equation}
where $\delta$ is initial overdensity, $\bar{\rho}_{a}(z_{\rm eq})$ is the average density of axions at matter-radiation equity. The overdensity distribution can be described approximately by \footnote{The form of overdensity distribution and the factors $a_{1}$ and $a_{2}$ are slightly different from Ref.~\cite{Fairbairn:2017sil}. The expression used here gives a very good fit to the curve for $\eta=4$ in Figure 2 in Ref.~\cite{Kolb:1995bu}. },
\begin{equation}\label{eq:MCdis}
\mathcal{F}(\delta>\delta_{0})=\frac{1}{\left[1+\left(\frac{\delta_{0}}{a_{1}}\right)^{2}\right]^{a_{2}}}\, ,
\end{equation}
By fitting with the curve for $\eta=4$ in Figure 2 in Ref.~\cite{Kolb:1995bu} we obtain the parameter values $a_{1}=1.004$ and $a_{2}=0.461$.

The density profile of axion MCs may depend on quantum pressure and self-interaction. In the following we neglect axion self-interaction. When quantum pressure becomes strong enough to balance gravity, an equilibrium configuration may form, called the axion star. The characteristic radius of dilute axion stars
is related to the axion mass~\cite{Schiappacasse:2017ham} \footnote{Here we use the characteristic radius $R$ defined in an axion star profile $\varphi(r)=\alpha_{0}{\rm sech}(r/R)$ with density $\rho(r)\propto \varphi(r)^{2}$. The definition gives a radius larger than the radius $r_{\rm sol}$ used in Appendix D in Ref.~\cite{Fairbairn:2017sil}.},
\begin{equation}\label{eq:axstar}
R=0.224\, {\rm pc} \left(\frac{M}{M_{\odot}}\right)^{-1}\left(\frac{m_{a}}{10^{-16}\,{\rm eV}}\right)^{-2}\, ,
\end{equation}
where $M$ is the mass of axion star. When the radius of MC or MCH is comparable to $R$, quantum pressure becomes important and the density is close to a constant near the center of MC or MCH. If the scale radius $r_{s}$ of a MC is much larger than the axion star radius $R$, the formation of the MC is pure gravitational, and the profile Eq.~(\ref{eq:NFW}) is still a good approximation. 

The condition $r_{s}\gg R$ is satisfied by MCHs discussed here. For dense MCHs with $c=1.4\times 10^{4}$, the scale radius is

\begin{equation}
r_{s}=1.51\times 10^{-3}\,{\rm pc}\left(\frac{m_{s}}{M_{\odot}}\right)^{\frac{1}{3}}\, .
\end{equation}
Take $m_{s}=M$ and use Eq.~(\ref{eq:M0}), the condition $r_{s}\gg R$ becomes $m_{s}\gg 0.2m_{0}$, independent of the axion mass. The binary disruption is dominated by cases where $m_{s}\gtrsim 10^{3}m_{0}$, which clearly satisfy $r_{s}\gg R$. The radius $r_{s}$ for diffuse MCHs are even larger. Hence it is reasonable to use NFW profile for dense and diffuse MCH models

For isolated MCs, the condition $r_{s}\gg R$ is not necessarily satisfied. By equating $\rho_{\rm mc}$ to the average density $m_{0}/(4\pi R^{3}/3)$, and using Eq.~(\ref{eq:M0}), we find $r_{s}\gg R$ is satisfied for initial overdensity $\delta\lesssim 20$. Because the MC density Eq.~(\ref{eq:MCd}) is derived from pure gravitational collapse, it can only be used for $\delta\lesssim 20$ where quantum pressure from ultralight particles can be neglected. For larger initial overdensity, the density of MCs does not increase for fixed MC mass $m_{0}$, and the profile of MCs becomes flat near center.

The radius $r_{s}$ of these MCs is fixed once the concentration $c$ is known. For a MC with initial overdensity $\delta$, the density $\rho_{\rm mc}(\delta)$ is obtained from Eq.~(\ref{eq:MCd}).  By equating $\rho_{\rm mc}(\delta)=\rho_{c}\delta_{\rm char}$ and using Eq.~(\ref{eq:cdelta}) we obtain the concentration $c(\delta)$ of the MC. For isolated MC case, we will consider all MC with $10^{-3}\lesssim\delta\lesssim 20$~\cite{Kolb:1994fi}.

For dense MCH case, tidal disruption within MCHs may disrupt MCs with a small concentration $\delta\lesssim 1$~\cite{Fairbairn:2017sil}. Hence we only consider those MCs with $\delta\gtrsim 1$ that group together to form MCHs. Note all MCs are assumed to have the same mass $m_{0}$.
The average density of these MCs is
\begin{equation}\label{eq:dense}
\rho_{\rm ave}=\frac{1}{\sum_{i}1/\rho_{{\rm mc},i}(\delta_{i})}\, .
\end{equation}

For a randomly sampled initial overdensity $\delta$ that satisfies the distribution described by Eq.~(\ref{eq:MCdis}), and only select those with $\delta>1$, the average density $\rho_{\rm ave}$ can be obtained. Then using $\rho_{\rm ave}=\rho_{c}\delta_{\rm char}$ and Eq.~(\ref{eq:cdelta}), we obtain the concentration $c_{\rm ave}=1.4\times 10^{4}$ for dense MCH model. The concentration $c_{\rm ave}$ only changes slightly if we further require $\delta\lesssim 20$. The MCH mass in dense MCH model that is relevant for binary star evaporation is larger than about $10^{3}m_{0}$, which means 
the summation in Eq.~(\ref{eq:dense}) includes at least $10^{3}$ MCs.
Hence the average density $\rho_{\rm ave}$ is a suitable estimation for dense MCH density. For dense MCH model, we use the concentration
$c_{\rm ave}=1.4\times 10^{4}$ for all MCHs. For diffuse MCHs, the concentration is taken from Ref.~\cite{Fairbairn:2017sil}.

In Fig.~\ref{img2} we plot the MCH concentration used here for diffuse MCHs, dense MCHs and simulations of Ref.~\cite{Xiao:2021nkb}. The concentration for dense MCHs is the largest, while for diffuse MCHs and the simulation results, the concentration is smaller and decreases for larger MCH mass with $m_{s}>m_{0}$.

To be more specific, we illustrate the characteristic size of MCs and MCHs here. For $m_{a}=10^{-14}\,\rm eV$ and $m_{s}=10^{6}m_{0}$, the diffuse MCHs have a concentration $c\sim 10$ while $c=1.4\times 10^{4}$ for dense MCHs. Using 
Eq.~(\ref{eq:rv}) and $c=r_{v}/r_{s}$, we find the scale radius $r_{s}\sim 180\,{\rm pc}$ for diffuse MCHs and $r_{s}\sim 0.13\,{\rm pc}$ for dense MCHs. For MCs with $\delta=1$, the concentration $c\sim 10^{4}$. The corresponding radius of MC is $r_{s}\sim 2\times 10^{-3}\,\rm pc$. Hence in general, the size of MCHs is much smaller than the Milky-Way, and the size of individual MCs is much smaller than MCHs.

\section{Binary star evaporation}
\label{sec:b}
As we have discussed, a fluctuation of gravitational potential tends to increase the orbital energy of the binary star. The density fluctuation is, 
\begin{equation}
\delta\rho(\vec{r},t)=\sum_{i}\rho_{i}(\vec{r}_{i},t)-\rho_{0}\, ,
\end{equation}
where we sum over individual MCHs labeled $i$. The fluctuation of gravitational potential can be described by a correlation function,
\begin{equation}\label{eq:corr}
\langle\delta\rho(\vec{r},t)\delta\rho(\vec{r}^{\,\prime},t')\rangle\equiv C_{\rho}(\vec{r}-\vec{r}^{\,\prime},t-t')\, .
\end{equation}

The Fourier transformation of the density correlation function is
\begin{equation}
C_{\rho}(\vec{k},\omega)=\int{\rm d}^{3} \vec{r} {\rm d}t \, C_{\rho}(\vec{r},t)e^{-i(\vec{k}\cdot\vec{r}-\omega t)}\, .
\end{equation}

When considering the binary star evaporation rate, we must average over dark matter perturbations during a time $T$.
This can be done when $T$ is much larger than the timescale of dark matter perturbations. Assuming that the dark matter density fluctuation has a characteristic length scale $\lambda_{\rm DM}$. To ensure we average over many dark matter perturbations, $T$ should be much larger than $\lambda_{\rm DM}/v$, where $v$ is the velocity of the stars relative to the dark matter background.

For wide binaries considered here, the orbital motion is slow. $T$ should be much smaller than the orbital period $2\pi/\omega_{b}$ so that the relative motion of binary stars can be neglected within $T$.
Hence we require
\begin{equation}\label{eq:condition}
\frac{\lambda_{\rm DM}}{v}\ll T\ll \frac{2\pi}{\omega_{b}}\, .
\end{equation}
As we will see below, the condition in Eq.~(\ref{eq:condition}) provides an upper limit for MCH mass, below which our calculation of evaporation remains valid.

The orbital energy growth rate is~\cite{Qiu:2024muo}
\begin{equation}\label{eq:E1}
\frac{\langle\Delta E\rangle}{T}=\mu\int\frac{\vec{k}^{2}{\rm d}^{3}k}{(2\pi)^{3}}C_{\Phi}\left(\vec{k},\vec{k}\cdot\vec{v}_{c}\right)\left\{1-\cos\left[\vec{k}\cdot(\vec{r}_{1}-\vec{r}_{2})\right]\right\}\, , 
\end{equation}
where $\mu$ is the reduced mass of the binary star, $\vec{v}_{c}$ is the center of mass velocity relative to the dark matter background, $C_{\Phi}(\vec{k},\omega)$ is the correlation function of gravitational potential, $C_{\Phi}=16\pi^{2}G^{2}k^{-4}C_{\rho}$.
We also assume a Maxwellian distribution of MCH velocity $\vec{v}$,
\begin{equation}\label{eq:DF}
F(\vec{v})=\frac{\rho_{0}}{(2\pi\sigma^{2})^{\frac{3}{2}}}e^{-\frac{v^{2}}{2\sigma^{2}}}  \, .
\end{equation}
where $\sigma$ is the standard deviation of soliton velocity.

From Eq.~(\ref{eq:corr}), the density correlation function for $N$ MCHs within volume $V$ is

\begin{equation}
\begin{aligned}
&C_{\rho}(\vec{r},t)=\frac{1}{(\rho_{0}V)^{N}}\int d^{3}\vec{r}_{1}d^{3}\vec{v}_{1}\ldots d^{3}\vec{r}_{N}d^{3}\vec{v}_{N} \\ &\left(\sum_{i}\rho_{i}(\vec{r_{i}})-\rho_{0}\right)\left(\sum_{j}\rho_{j}(\vec{r}-\vec{r_{j}}-\vec{v_{j}}t)-\rho_{0}\right)\\ & F(\vec{v}_{1})\ldots F(\vec{v}_{N}) \, ,
\end{aligned}
\end{equation}
We first multiply terms in the two brackets.
The integral receives constant contributions for terms involving one or two factors of $\rho_{0}$ and terms with $i\neq j$. These terms only contribute a zero component to $C_{\rho}(\vec{k},\omega)$ and do not affect binary evaporation. Keeping terms with $i=j$ yields

\begin{equation}
C_{\rho}(\vec{r},t)=\frac{1}{\rho_{0}V}\sum_{i} \int {\rm d}^{3} \vec{v}_{i}\, {\rm d}^{3} \vec{r}_{i} \rho_{i}(\vec{r}_{i})\rho_{i}(\vec{r}-\vec{r}_{i}-\vec{v}_{i}t)F(\vec{v}_{i})   \, .
\end{equation}
where we sum over individual MCHs labeled $i$ in a volume $V$. After a Fourier transformation and changing integration variable, we obtain
\begin{equation}
\begin{aligned}
C_{\rho}(\vec{k},\omega)&=\frac{1}{\rho_{0}V}\sum_{i}\int 
{\rm d}^{3} \vec{r}_{i}\, {\rm d}^{3} \vec{r}^{\,\prime}_{i}\, {\rm d}^{3} \vec{v}_{i}\, {\rm d} t\, \\
&\rho_{i}(\vec{r}_{i})\rho_{i}(\vec{r}^{\,\prime}_{i})F(\vec{v}_{i})e^{-i\vec{k}\cdot(\vec{r}_{i}+\vec{r}^{\,\prime}_{i}+\vec{v}_{i}t)}e^{i\omega t}   \, .
\end{aligned}
\end{equation}

Using Eq.~(\ref{eq:DF}) and integrate over $\vec{v}_{i}$ and $t$ yields
\begin{equation}
C_{\rho}(\vec{k},\omega)=\frac{1}{V}\sum \rho_{m}^{2}(\vec{k})\sqrt{\frac{2\pi}{k^{2}\sigma^{2}}}e^{-\frac{\omega^{2}}{2k^{2}\sigma^{2}}}\, .
\end{equation}
Here we sum over different MCHs inside $V$. $\rho_{m}(\vec{k})$ is the Fourier transformation of MCH density profile, which depends on MCH mass $m_{s}$,
\begin{equation}
\rho_{m}(\vec{k})=\int {\rm d}^{3} \vec{r}\, \rho_{m}(\vec{r})e^{-i\vec{k}\cdot\vec{r}}\, .
\end{equation}
In particular, for MCH profile Eq.~(\ref{eq:NFW}), 
\begin{equation}
\begin{aligned}
\rho_{m}(\vec{k})&=2\pi\rho_{s}r_{s}^{3}\\ &\left[-2\cos(kr_{s}){\rm Ci}(kr_{s})+\sin(kr_{s})(\pi-2{\rm Si}(kr_{s}))\right]\, ,
\end{aligned}
\end{equation}
where $k=|\vec{k}|$, $\rho_{s}\equiv \rho_{c}\delta_{\rm char}$ is the characteristic density of MCHs. Si and Ci are sine integral and cosine integral, respectively.

For MCHs with a mass function ${\rm d}n/{\rm d}\ln m_{s}$, the density correlation function becomes
\begin{equation}
C_{\rho}(\vec{k},\omega)=\left[\int\frac{{\rm d}n}{{\rm d}\ln m_{s}} \frac{\rho_{m}^{2}(\vec{k})}{m_{s}} {\rm d}m_{s}\right]\sqrt{\frac{2\pi}{k^{2}\sigma^{2}}}e^{-\frac{\omega^{2}}{2k^{2}\sigma^{2}}}\, ,
\end{equation}
Using Eq.~(\ref{eq:E1}) and $C_{\Phi}=16\pi^{2}G^{2}k^{-4}C_{\rho}$,
the orbital energy growth rate can be written as
\begin{equation}
\frac{\langle\Delta E\rangle}{T}=\left(\frac{\langle\Delta E\rangle}{T}\right)_{0}A\, .
\end{equation}
The $(\langle\Delta E\rangle/T)_{0}$ has the same form as the result for binary disruption by stars,
\begin{equation}
\left(\frac{\langle\Delta E\rangle}{T}\right)_{0}=\frac{8\pi\mu\rho_{0}G^{2}m_{0}}{v_{c}} \, .
\end{equation}
The dimensionless factor $A$ is an integration over different MCH mass weighted by MCH mass function,
\begin{equation}\label{eq:A}
A(a)=\frac{1}{\sqrt{2\pi}}\frac{v_{c}}{\sigma}\int \left(\frac{m_{s}}{m_{0}}\right)^{-2}\frac{m_{s}}{\rho_{0}}\frac{{\rm d}n}{{\rm d}\ln m_{s}}A_{0}\left(\frac{r_{s}}{a}\right){\rm d}\left(\frac{m_{s}}{m_{0}}\right)\, .
\end{equation}
For a given MCH mass $m_{s}$, the corresponding $c(m_{s})$ and $r_{s}(m_{s})$ can be determined in each model. 

We use a coordinate system in which $+z$ axis is parallel to $\vec{v}_{c}$. The relative position of the two stars is $\vec{r}_{1}-\vec{r}_{2}=(r_{x},r_{y},r_{z})$. Then the factor $A_{0}$ is~\cite{Qiu:2024muo}

\begin{equation}\label{eq:A0}
\begin{aligned}
A_{0}\left(\frac{r_{s}}{a}\right)&=\int_{0}^{+\infty}\frac{{\rm d}k}{k}\frac{\rho_{m}^{2}(k)}{m_{0}^{2}}\int_{-1}^{1}{\rm d}x\, e^{-\frac{v_{c}^{2}x^{2}}{2\sigma^{2}}}\\ &\left[1-J_{0}\left(k\sqrt{r_{x}^{2}+r_{y}^{2}}\sqrt{1-x^{2}}\right)\cos(kr_{z}x)\right]\, .
\end{aligned}
\end{equation}

\begin{figure}[htbp]
  \centering
  \includegraphics[width=9cm]{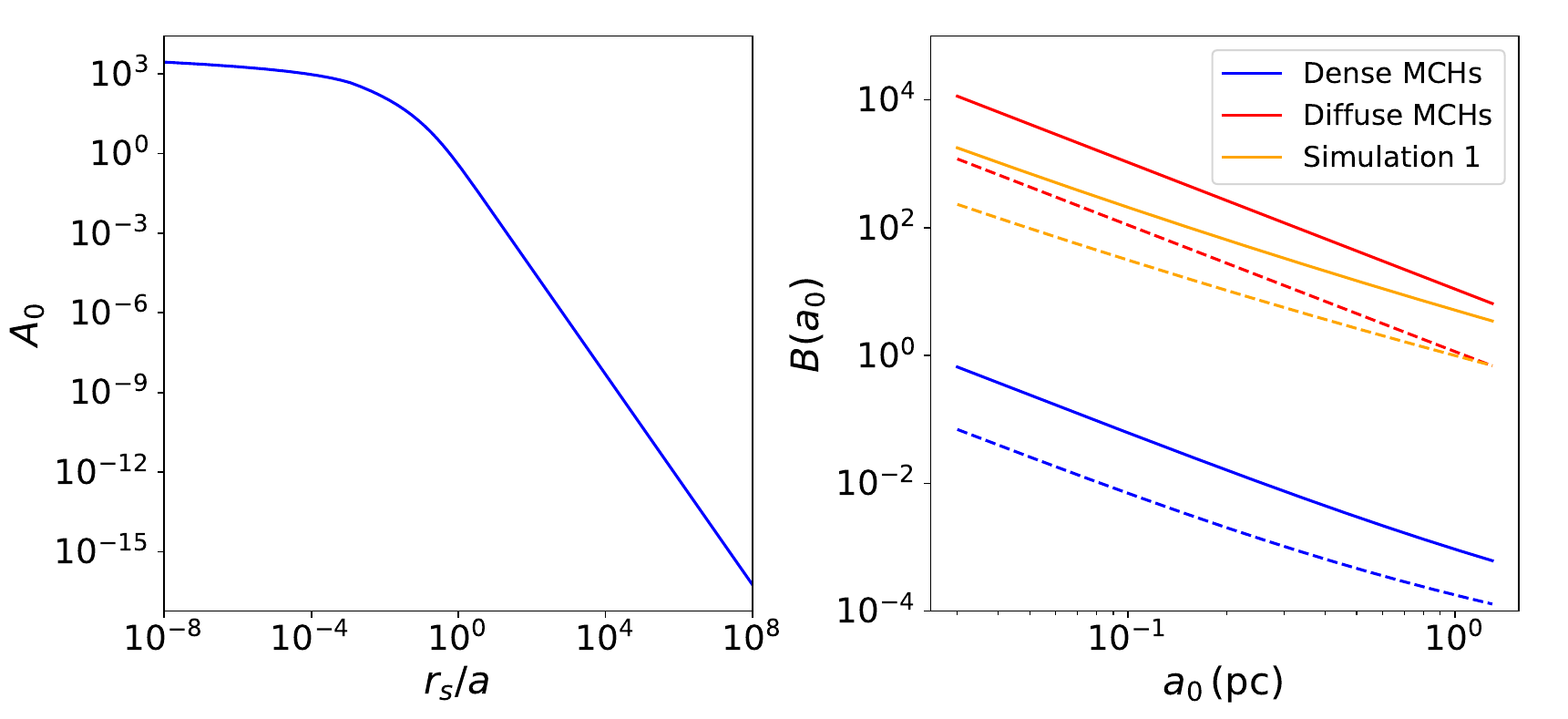}\\
  \caption{\emph{Left}: The correction factor $A_{0}(r_{s}/a)$. $A_{0}$ 
 is independent of MCH models. \emph{Right}: The correction factor $B(a_{0})$ for dense MCHs, diffuse MCHs and simulation 1. Solid lines represent $m_{a}=10^{-16}\,{\rm eV}$ while dashed lines represent $m_{a}=10^{-15}\,{\rm eV}$. }
  \label{img3}
\end{figure}

We define the inclination angle $\alpha$ as the angle between $\vec{v}_{c}$ and the normal vector of the orbital plane. The factor $A_{0}$ only slightly depend on $\alpha$. We will average over the orbital orientation and orbital relative position to obtain the averaged evaporation rate.  We also take $v_{c}\approx\sigma$ throughout the paper. After these orbital averages, $A_0$ only depends on $r_{s}/a$. In the left panel of Fig.~\ref{img3} we plot the factor $A_{0}\left(r_{s}/a\right)$. For large MCH radius $r_{s}/a\gg 1$, the factor $A_{0}$ has a power-law dependence on $r_{s}/a$. While for small MCH radius, the factor $A_{0}$ depends logarithmic on $r_{s}/a$.

When the total (kinetic and potential) orbital energy $E_{T}$ increase to zero, the binary star is completely disrupted. The disruption time for an initial binary star separation $a_{0}$ is
\begin{equation}\label{eq:disruptT5}
\begin{aligned}
t_{d}&=14.3\, {\rm Gyr}\left(\frac{M_{T}}{0.5\,M_{\odot}}\right) \left(\frac{a_{0}}{0.1\,\rm pc}\right)^{-1}\left(\frac{v_{c}}{200\,\rm {km/s}}\right)\\ &\left(\frac{m_{s}}{30\,M_{\odot}}\right)^{-1}\left(\frac{\rho_{0}}{0.4\,\rm {GeV/cm^{3}}}\right)^{-1}B\left(a_{0}\right)\, ,
\end{aligned}
\end{equation}
where the correction factor $B(a_{0})$ is given by
\begin{equation}
B(a_{0})=\int_{0}^{1} {\rm d}u \left/ A\left(
\frac{a_{0}}{u}\right) \right. .
\end{equation}

Note that $B(a_{0})$ depends on the MCH mass function and concentration. Hence we obtain different values of $B(a_{0})$ for different MCH models. In the right panel of Fig.~\ref{img3} we plot $B(a_{0})$ for diffuse MCHs, dense MCHs and simulation 1. For diffuse and dense MCHs, PS formalism is used. Solid lines represent $m_{a}=10^{-16}\,{\rm eV}$ while dashed lines represent $m_{a}=10^{-15}\,{\rm eV}$. Dense MCH model corresponds to a much smaller $B(a_{0})$, which means binary star disruption easily occurs.  

In the isolated MC model, the mass of MCs are $m_{0}$ while their radius depends on overdensity $\delta$. Here the above calculation of binary evaporation rate is only a little bit different. In Eq.~(\ref{eq:A0}), $\rho_{m}(k)$ should be changed to $\rho_{\delta}(k)$ because the density profile depends on $\delta$. Eq.~(\ref{eq:A}) should be changed to
\begin{equation}\label{eq:Ad}
A(a)=\frac{1}{\sqrt{2\pi}}\frac{v_{c}}{\sigma}\int \frac{m_{0}}{\rho_{0}}A_{0}\left(\frac{r_{s}}{a}\right)\frac{{\rm d}n}{{\rm d}\delta}{\rm d}\delta\, ,
\end{equation}
where ${\rm d}n/{\rm d}\delta$ is given by Eq.~(\ref{eq:MCdis}) and the normalization condition becomes
\begin{equation}\label{eq:Ad}
\int \frac{m_{0}}{\rho_{0}}\frac{{\rm d}n}{{\rm d}\delta}{\rm d}\delta=1\, ,
\end{equation}
where we integrate all $\delta$ with $10^{-3}\lesssim\delta\lesssim 20$. The calculation of factor $B$ is unchanged.

\section{Implication for ultralight axions}
\label{sec:c}
In what follows we will use GAIA's binary star data as presented by Catalog I and II in Ref.~\cite{Qiu:2024muo} to constrain axion MCH models. The two catalogs include `halo-like' binary stars with projected separation larger than $0.3\,\rm pc$. To be halo-like, these candidates have a large velocity perpendicular to galactic plane, hence during the majority time these binary stars stay away from the disk, and they are more perturbed by dark matter in the halo.

\begin{figure}[t]
  \centering
  \includegraphics[width=8cm]{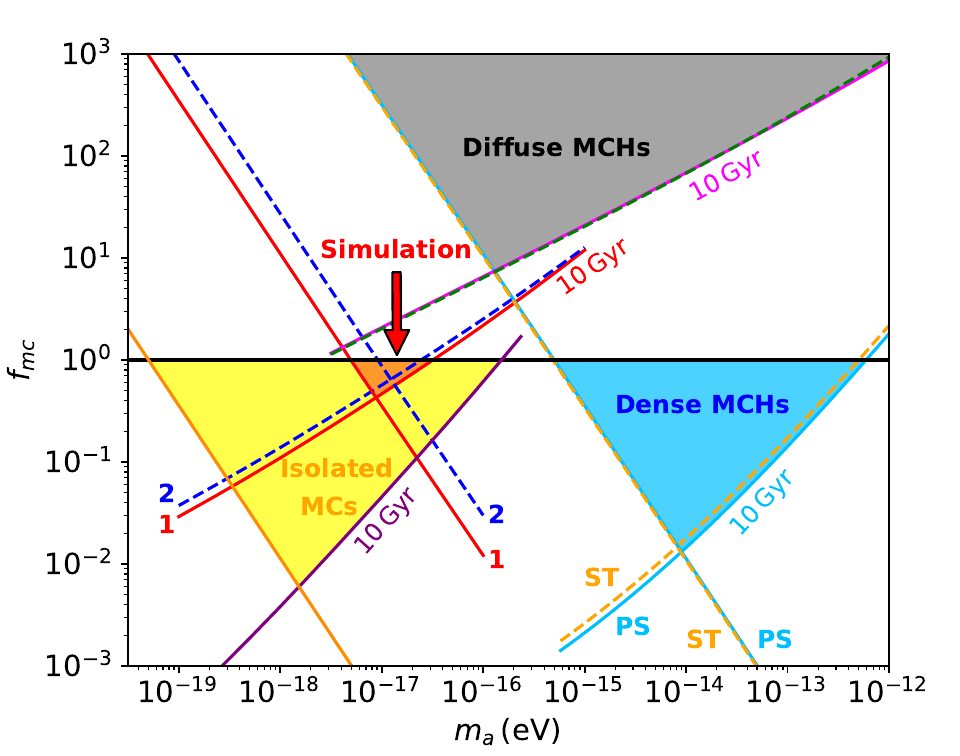}\\
  \caption{Constraints on axion mass and the fraction of MCH mass $f_{\rm mc}$ for different MCH models. The blue, red, and yellow shaded regions are constraints for dense MCH model, MCHs from simulation 1 and isolated MC model. For diffuse MCH model, the gray shaded region has $f_{\rm mc}>1$, meaning diffuse MCHs are not constrained by binary disruption. We mark the constraints for PS and ST formalism and the constraints from binary evaporation assuming an average lifetime $10\,\rm Gyr$.}
  \label{img4}
\end{figure}

First we need to check the theoretical validity conditions of our binary evaporation rate. We have assumed an average binary disruption time to be longer than $10\,{\rm Gyr}$, and also assumed the orientation of the binary stars to be statistically randomized. Hence the physical separation of binary stars $a=4a_{\perp}/\pi$, where $a_{\perp}$ is the separation projected to a plane perpendicular to the line of sight.
The constraints for different MCH models are shown in Fig.~\ref{img4}. As wide binary systems' orbit period is typically very long, our evaporation rate calculation assumes a slow orbit approximation, requiring Eq.~(\ref{eq:condition}) to be satisfied. Slow orbit approximation requires an upper bound of the MCH mass. To see this, we take binary star velocity $v_{c}=200\,{\rm km/s}$. The mean MCH separation is then $\lambda_{\rm DM}=(f_{\rm mc}\rho_{\rm DM}/m_{s})^{-1/3}$. The average orbital period for binaries in Catalog I and II is $2\pi/\omega_{b}=1.26\times 10^{15}\,\rm s$. Thus, our calculation validity condition $\lambda_{\rm DM}/v_{c}<2\pi/\omega_{b}$ requires an upper bound on the MCH mass:
\begin{equation}\label{eq:upperM0}
m_{s}<5.67\times 10^{9}\,M_{\odot}\cdot f_{\rm mc}  \, .
\end{equation}
In addition, we would also require the mean separation of MCHs are less than the galactic scale $\mathcal{O}(\rm kpc)$, namely MCHs are not too scarce so that we may take the same statistical distribution of dark matter density over a binary's trajectory. $\lambda_{\rm DM}<1\,\rm kpc$ gives a tighter requirement:
\begin{equation}\label{eq:upperM0}
m_{s}<10^{7}\,M_{\odot}\cdot f_{\rm mc}  \, ,
\end{equation}
and we assume $90\%$ of the MCH masses in each model satisfy this condition. To be more specific, we introduce a characteristic MCH mass $m_{\rm max}$ defined as
\begin{equation}\label{eq:90mass}
\int_{0}^{m_{\rm max}} \frac{{\rm d}n}{{\rm d}\ln m_{s}} {\rm d}m_{s}=0.9\rho_{0} \, ,
\end{equation}
in which the ratio $m_{\rm max}/m_{0}$ depends on the MCH model, and can be obtained for each model. Hence Eq.~(\ref{eq:upperM0}) becomes
\begin{equation}\label{eq:upperM}
m_{\rm max}<10^{7}\,M_{\odot}\cdot f_{\rm mc}  \, ,
\end{equation} 
and we will use it as the main validity constraint, yielding the left-boundary in each shaded region in Fig.~\ref{img4}. 
The other boundary on the right (larger $m_a$) side of each shaded region in Fig.~\ref{img4} corresponds to average disruption time of $t_d\geq 10\,{\rm Gyr}$ for the wide binaries in Catalog I and II.

The constraint for the diffuse MCH model is shown by the gray shaded region, which completely falls into the unphysical $f_{\rm mc}>1$ area. Hence the disruption of Milky Way's wide binaries does not constrain this model. In comparison, overdensities are denser with dense MCHs, isolated MCs and simulated mass functions, and binary disruption does yield constrained regions below $f_{\rm mc}=1$. We can see from Fig.~\ref{img2} that the concentration from simulations appear to be smaller than that in diffuse MCH models for MCHs with the same $m_{s}/m_{0}$. However, as seen in Fig.~\ref{img1}, the MCH mass function from simulations concentrates on smaller $m_{s}/m_{0}$ side, yielding a larger concentration compared to diffuse MCH models. Hence the constraints for MCHs from simulations are tighter compared with diffuse MCH models. For dense MCH model, PS and ST formalisms lead to slightly different limits and they are represented by the blue solid and orange dashed lines, respectively. The constrained region for isolated MCs is shown by the yellow shaded region.

%The magenta solid line and the green dashed line are binary star constraints for diffuse MCH model with PS and ST formalism, which almost overlap in Fig.~\ref{img4}.  

Limits for the two simulated MCH mass functions from Ref.~\cite{Xiao:2021nkb} (denoted as Simulation 1 and 2) are shown by the shaded regions enclosed by orange and blue constraint curves, respectively, near the center of Fig.~\ref{img4}. Simulation 1 assumes the growth of axion minihalos
is terminated when the larger host halo is at a turnaround point ($\delta_{c}\approx 1.06$), while for simulation 2 it is later at the host halo's collapse ($\delta_{c}\approx 1.69$)~\cite{Xiao:2021nkb}. This difference lets simulation 2 to have a larger MCH mass, and an smaller overall MCH concentration, resulting in a slightly weaker tidal evaporation constraint.

\section{Discussions}
\label{sec:d}

To summarize, we considered binary star disruption by dark matter density fluctuations from MCs and MCHs. Considering the observed population of wide-separation `halo-like' binary system in GAIA data, the presence of these system can be interpreted into limits on the strength of short-distance tidal disruption from the MCs and MCHs.

We considered various MCH mass functions and concentrations, including diffuse MCH model, dense MCH model, isolated MC model
and simulated MCH distributions from literature. We also extended the calculations of binary evaporation rate to the case that MCs or MCHs have a range of mass and size. The binary star disruption time is calculated for each model. By assuming that the wide binaries samples to have a dark-matter tidal evaporation time longer than $10\,\rm Gyr$, we obtain the constraints for ALP mass and the fraction in the dark matter halo $f_{\rm mc}$. 

Wide binary disruption can place significant constraints for axion mass and MCH fraction $f_{\rm mc}$ for the models we considered, except for diffuse MCH model. For dense MCH model, isolated MC model and MCHs from simulations, the constraint parameter space covers the ultralight axion mass range around $10^{-15}-10^{-12}\,{\rm eV}$, $10^{-19}-10^{-16}\,{\rm eV}$ and $10^{-17}\,{\rm eV}$, respectively. 

In perspective, studying binary disruption by MCs and MCHs may provide further information for their mass function parameters $m_{0}$ and $A_{\rm osc}$. The MC characteristic mass $m_{0}$ is directly related to the axion mass at $T_{\rm osc}$, and $A_{\rm osc}$ is related to the initial power spectrum that potentially carries information from the PQ symmetry breaking process and the decay of topological defects.

\bigskip
{\bf Acknowledgements.}\\
The authors acknowledge support by the National Natural Science Foundation of China (No. 12275278 and 12150010).

\bibliography{refs}
\end{document}